\documentclass[12pt]{article}

\usepackage{epsfig,a4,amssymb}
\divide\textwidth\mag \multiply\textwidth by 1000
\divide\textheight\mag \multiply\textheight by 1000
\newcommand{\Real}{\mathop{\rm Re}\nolimits}
\newcommand{\Imag}{\mathop{\rm Im}\nolimits}
\newcommand{\ein}{{\rm e}}
\newcommand{\iim}{{\rm i}}
\newcommand{\sdt}{\hspace*{-1em}.\ \hangindent=1em\relax}
\makeatletter
\renewcommand{\@makecaption}[2]{%
	 \vskip 10\p@
	 \setbox\@tempboxa\hbox{#1. #2}%
	 \ifdim \wd\@tempboxa >\hsize
			 #1. #2\par
		 \else
			 \hbox to\hsize{\hfil\box\@tempboxa\hfil}%
	 \fi}
\renewcommand\maketitle{\par
 \begingroup
   \def\thefootnote{\fnsymbol{footnote}}%
   \def\@makefnmark{\hbox
       to\z@{$\m@th^{\@thefnmark}$\hss}}%
   \if@twocolumn
     \twocolumn[\@maketitle]%
     \else \newpage
     \global\@topnum\z@
     \@maketitle \fi\@thanks
 \endgroup
 \setcounter{footnote}{0}%
 \let\maketitle\relax
 \let\@maketitle\relax
 \gdef\@thanks{}\gdef\@author{}\gdef\@title{}\let\thanks\relax}
\renewcommand{\section}{\@startsection{section}{1}{0pt}%
{4.5ex plus 1ex minus 0ex}{3.3ex plus .2ex}{\Large\bf}}
\renewcommand{\subsection}{\@startsection{subsection}{2}{0pt}%
{4.5ex plus 1ex minus 0ex}{3.3ex plus .2ex}{\large\bf}}
\renewcommand{\@biblabel}[1]{#1.\ }
\renewcommand{\@oddhead}{\hfill \thepage}
\renewcommand{\@oddfoot}{\hfil \hfil}
\makeatother
\begin{document}

\sloppy
\raggedbottom

\thispagestyle{empty}

\title{\vspace*{-2.5em}\bf Quantum Electron Plasma, Visible and
Ultraviolet P-wave and Thin Metallic Film \vspace*{.5em}}

\author{A.A. Yushkanov$^1$ \ and \ N.V. Zverev$^2$ \vspace*{.5em}
\and
\sl Faculty of Physics and Mathematics, \\
\sl Moscow Region State University, \\
\sl Radio str. 10a, 100500 Moscow, Russia
}

\footnotetext[1]{\ \ yushkanov@inbox.ru}
\footnotetext[2]{\ \ zverev\_nv@mail.ru}

\date{}

\maketitle

\vspace*{-.5em}

\begin{abstract}
{\normalsize
The interaction of the visible and ultraviolet electromagnetic P-wave
with the thin flat metallic film localized between two dielectric media
is studied numerically in the framework of the quantum degenerate
electron plasma approach. The reflectance, transmittance and absorptance
power coefficients are chosen for investigation. It is shown that for
the frequencies in the visible and ultraviolet ranges, the quantum
power coefficients differ from the ones evaluated in framework of both
the classical spatial dispersion and the Drude -- Lorentz approaches.

\vspace*{.5em}


\vspace*{.5em}

}
\end{abstract}

\vspace*{1em}

\section{\sdt Introduction}

Studies of the electromagnetic waves interacting with tiny or nanoscale
metallic objects attract a large attention for a long time \cite{FuKlPa}
-- \cite{LaYu2}. Such investigations have not only theoretical interest
but are aimed also on multiple practical applications. In these
researches, the kinetic theory for the Fermi -- Dirac electron gas
is widely exploited. Using the theory, one takes into account the
spatial dispersion of the electron plasma. And moreover, the theory
of the electromagnetic waves interacting with flat metallic film was
well developed for the case of specular electron reflection
\cite{KlFu1,JoKlFu}.

But almost always in these researches, the quantum wave properties
of electrons in the electron plasma were disregarded. The main problem
was in a getting acceptable both the longitudinal and the transverse
dielectric functions (permittivities) of the quantum electron plasma
\cite{Lndh,Mrm}, \cite{LaYu3} -- \cite{LaYu6}. However since the
electrons in a metal obey in general the quantum laws, the quantum wave
electron effects should contribute to interaction of light with a metal.
It would be essential in case of the high frequency light and for the
nanoscale metallic objects.

In the present paper, we study the interaction of the visible and
ultraviolet electromagnetic P-wave with quantum degenerate electron
plasma in the thin flat metallic film localized between transparent
dielectric media. Being interested in the visible and ultraviolet light
we consider the frequencies of order and larger than the electron plasma
frequency. The values taken by us for investigation are the reflectance,
transmittance and absorptance power coefficients. We study the quantum
degenerate electron plasma with invariable relaxation time in case of
specular electron reflection from the film surface. The longitudinal
and transverse dielectric functions (permittivities) of the quantum
electron plasma are taken in the Mermin approach \cite{Mrm,LaYu5,LaYu6}.
We study the power coefficients as functions of frequency, of incidence
angle and of quantum parameter. We compare also the coefficients with
those obtained both in case of the Drude -- Lorentz theory without
spatial dispersion and in case of the classical degenerate electron
gas approach accounting for the spatial dispersion.

\section{\sdt The model and the power coefficients}

\begin{figure}[ht]

\vspace*{0mm}
\hspace*{15mm}
\epsfig{file=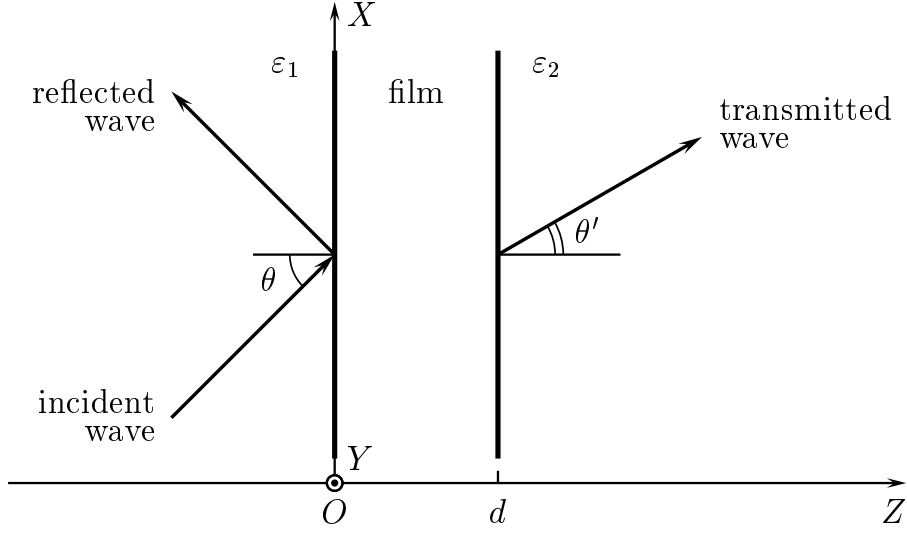,width=.75\textwidth}

\vspace*{0mm}

\caption{The metallic film between two dielectric media with
$\varepsilon_1$ and $\varepsilon_2$, and the incident,
reflected and transmitted waves.}
\label{fig:mt-flm}

\end{figure}

Let us consider the thin flat metallic film of the width $d$ placed
between two isotropic nonmagnetic transparent dielectric media with the
positive dielectric constants $\varepsilon_1$ and $\varepsilon_2$. So
the light dispersion and absorption of these media are disregarded by
us. We consider the electromagnetic wave is incident from the first
dielectric medium on the film under the angle $\theta$ from the surface
normal (see fig.~\ref{fig:mt-flm}). Hence the second medium can be
treated as a substrate.

We direct the $Z$ axis perpendicular to the film surface towards the
second dielectric medium. Let $z = 0$ be the surface contacting with
the first medium and therefore, the $z = d$ is the second surface having
contact with the second medium (fig.~\ref{fig:mt-flm}). We direct the
$X$ axis along the film surface in the incidence plane towards the wave
propagation.

We study the P-waves i.e. the ${\bf E}$ vectors of the waves incident
on, reflected from and transmitted through the film lie in the incidence
plane. We direct the third $Y$ axis in such a way that the rectangular
system being the right-handed one. Then the ${\bf H}$ vectors of the
waves are parallel to the $Y$ axis.

The electric and magnetic fields of the waves in the $z < 0$ space
in the projections onto $X$ and $Y$ axes look as follows
\cite{KlFu1}:
\begin{equation}\label{em_1}
\left\lbrace
\begin{array}{lll}
E_x(x,y,z,t) & = & \ein^{\iim(k_{1x}x -\omega t)}\bigl[a_I\,%
\ein^{\iim k_{1z}z} - a_R\,\ein^{-\iim k_{1z}z}\bigr]\cos\theta,
\\
H_y(x,y,z,t) & = & \displaystyle\ein^{\iim(k_{1x}x -\omega t)}%
\bigl[a_I\,\ein^{\iim k_{1z}z} + a_R\,\ein^{-\iim k_{1z}z}\bigr]%
\frac{\sqrt{\varepsilon_1}}{Z_0}.
\end{array}
\right.
\end{equation}

In the $z > d$ space, these fields look as
\begin{equation}\label{em_2}
\left\lbrace
\begin{array}{lll}
E_x(x,y,z,t) & = & \ein^{\iim(k_{2x}x -\omega t)}%
a_T\cos\theta'\,\ein^{\iim k_{2z}(z - d)},
\\
H_y(x,y,z,t) & = & \displaystyle\ein^{\iim(k_{2x}x -\omega t)}%
a_T\frac{\sqrt{\varepsilon_2}}{Z_0}\,\ein^{\iim k_{2z}(z - d)}.
\end{array}
\right.
\end{equation}

Here $\omega$ is the wave frequency, $k_{1x}$ and $k_{1z}$ are the
$x$- and $z$-coordinates of the incident wave vector ${\bf k}_1$ in
the first dielectric medium, $k_{2x}$ and $k_{2z}$ are the same
coordinates of the transmitted wave vector ${\bf k}_2$ in the second
dielectric medium:
\begin{eqnarray}\label{kx}
k_{1x} = \frac{\omega}{c}\sqrt{\varepsilon_1}\sin\theta = k_{2x} =
\frac{\omega}{c}\sqrt{\varepsilon_2}\sin\theta';
\\
k_{1z} = \frac{\omega}{c}\sqrt{\varepsilon_1}\cos\theta, \qquad
k_{2z} = \frac{\omega}{c}\sqrt{\varepsilon_2}\cos\theta'.
\nonumber
\end{eqnarray}
Further, the $a_R$, $a_I$ and $a_T$ denote respective complex electric
field amplitudes of the incident, reflected and transmitted waves. The
$c$ is the vacuum speed of light, $Z_0$ denotes the dimensional (in Ohm)
vacuum impedance. And at the end, $\theta'$ is the narrow refraction
angle into the second dielectric medium from the surface normal
(fig.~\ref{fig:mt-flm}). Note that in the case of total internal
reflection \ $\sin\theta' > 1$, \ the $\cos\theta'$ value is pure
imaginary with the \ $\Imag\cos\theta' > 0$. So, the $\cos\theta'$ is
evaluated according to the prescription following from (\ref{kx}):
\begin{equation}\label{csthpr}
\cos\theta' = \left\lbrace
\begin{array}{lll}
\displaystyle \sqrt{1 - \frac{\varepsilon_1}{\varepsilon_2}%
\sin^2\theta}, & & \displaystyle \sin\theta \leqslant
\sqrt{\frac{\varepsilon_2}{\varepsilon_1}};
\\[1em]
\displaystyle \iim\sqrt{\frac{\varepsilon_1}{\varepsilon_2}%
\sin^2\theta - 1}, & & \displaystyle \sin\theta >
\sqrt{\frac{\varepsilon_2}{\varepsilon_1}}.
\end{array}
\right.
\end{equation}

In the film space $0 < z < d$, the electric and magnetic fields may be
represented in the following manner:
\begin{equation}\label{em_sl}
\left\lbrace
\begin{array}{lll}
E_x(x,y,z,t) & = & \ein^{\iim(k_{1x}x -\omega t)}\bigl[\alpha_1
E^{(1)}_x(z) + \alpha_2 E^{(2)}_x(z)\bigr],
\\
H_y(x,y,z,t) & = & \ein^{\iim(k_{1x}x -\omega t)}\bigl[\alpha_1
H^{(1)}_y(z) + \alpha_2 H^{(2)}_y(z)\bigr].
\end{array}
\right.
\end{equation}
Here $\alpha_1$ and $\alpha_2$ are some constant coefficients, and
$E^{(j)}_x(z)$ and $H^{(j)}_y(z)$ are the symmetric or antisymmetric
modes of electric and magnetic fields ($j = 1,2$):
\begin{equation}\label{em_md}
E^{(j)}_x(z) = (-1)^j E^{(j)}_x(d - z), \qquad
H^{(j)}_y(z) = (-1)^{j+1} H^{(j)}_y(d - z).
\end{equation}

On the film surfaces $z = 0$ and $z = d$, one has the boundary
conditions:
\begin{equation}\label{em_bc1}
\left\lbrace
\begin{array}{lll}
E_x(x,y,-0,t) & = & E_x(x,y,+0,t),
\\
H_y(x,y,-0,t) & = & H_y(x,y,+0,t);
\end{array}
\right.
\end{equation}
\begin{equation}\label{em_bc2}
\left\lbrace
\begin{array}{lll}
E_x(x,y,d-0,t) & = & E_x(x,y,d+0,t),
\\
H_y(x,y,d-0,t) & = & H_y(x,y,d+0,t).
\end{array}
\right.
\end{equation}

Following \cite{KlFu1} -- \cite{KlFu2}, we consider the dimensionless
surface inpedance for the P-wave modes on the $z = 0$ surface
($j = 1,2$):
\begin{equation}\label{sf_imp1}
Z^{(j)}_P = \frac{1}{Z_0}\frac{E^{(j)}_x(+0)}{H^{(j)}_y(+0)}.
\end{equation}

Substituting (\ref{em_1}) and (\ref{em_sl}) into (\ref{em_bc1}),
(\ref{em_2}) and (\ref{em_sl}) into (\ref{em_bc2}), taking into
account the property (\ref{em_md}) of the modes and using the surface
impedance (\ref{sf_imp1}), after elimination of similar multipliers
one gets the following system:
\begin{equation}\label{em_sys1}
\left\lbrace
\begin{array}{rcl}
(a_I - a_R)\cos\theta & = & \alpha_1 E^{(1)}_x(+0) +
\alpha_2 E^{(2)}_x(+0),
\\
(a_I + a_R)\sqrt{\varepsilon_1} & = & \displaystyle
\alpha_1\frac{E^{(1)}_x(+0)}{Z^{(1)}_P} +
\alpha_2\frac{E^{(2)}_x(+0)}{Z^{(2)}_P},
\\[1em]
a_T\cos\theta' & = & -\alpha_1 E^{(1)}_x(+0) +
\alpha_2 E^{(2)}_x(+0),
\\
a_T\sqrt{\varepsilon_2} & = & \displaystyle
\alpha_1\frac{E^{(1)}_x(+0)}{Z^{(1)}_P} -
\alpha_2\frac{E^{(2)}_x(+0)}{Z^{(2)}_P}.
\end{array}
\right.
\end{equation}

After elimination of the values $\alpha_j E^{(j)}_x(+0)$ ($j = 1,2$)
from the system (\ref{em_sys1}), one comes to the system
\begin{equation}\label{em_sys2}
\left\lbrace
\begin{array}{rcl}
a_T & = & U^{(1)}_P a_I - V^{(1)}_P a_R,
\\
a_T & = & -U^{(2)}_P a_I + V^{(2)}_P a_R.
\end{array}
\right.
\end{equation}
Here we denoted ($j = 1,2$):
\begin{equation}\label{uv_c}
U^{(j)}_P = \frac{\cos\theta - Z^{(j)}_P\sqrt{\varepsilon_1}}%
{\cos\theta' + Z^{(j)}_P\sqrt{\varepsilon_2}}, \qquad
V^{(j)}_P = \frac{\cos\theta + Z^{(j)}_P\sqrt{\varepsilon_1}}%
{\cos\theta' + Z^{(j)}_P\sqrt{\varepsilon_2}}.
\end{equation}

\bigskip

Now we turn to the definition of the reflectance $R$, transmittance $T$
and absorptance $A$ power coefficients. The first two of them are merely
the following ratios \cite{LdLf,DrGr}:
\begin{equation}\label{rt_def}
R = \frac{|\langle S_z\rangle_R|}{|\langle S_z\rangle_I|}, \qquad
T = \frac{|\langle S_z\rangle_T|}{|\langle S_z\rangle_I|}.
\end{equation}
Here $\langle S_z\rangle$ is the time averaged energy flux density, or
Poynting, vector, projected onto the $Z$ axis:
\begin{equation}\label{Pnt1}
\langle S_z\rangle = \frac{1}{2}\Real({\bf E}\times{\bf H}^*)\cdot
{\bf e}_z,
\end{equation}
where $^*$ denotes the complex conjugation and ${\bf e}_z$ is the
unit vector towards the $Z$ axis direction. In the equations
(\ref{rt_def}), the $I$, $R$ and $T$ subscript letters stand for the
incident, reflected and transmitted waves respectively. For the P-wave,
the expression (\ref{Pnt1}) can be transformed to the following one:
\begin{equation}\label{Pnt2}
\langle S_z\rangle = \frac{1}{2}\Real(E_x H^*_y).
\end{equation}

We substitute the expressions (\ref{em_1}) and (\ref{em_2}) to the
(\ref{Pnt2}) and select the terms related to $\langle S_z\rangle_I$,
$\langle S_z\rangle_R$ and $\langle S_z\rangle_T$. Then we substitute
these terms to the equations (\ref{rt_def}) and take into account the
positivity of the dielectric constants $\varepsilon_1$ and
$\varepsilon_2$. And one arrives at the following expressions
for the reflectance $R$ and transmittance $T$ power coefficients:
\begin{equation}\label{rt_cf1}
R = \left|\frac{a_R}{a_I}\right|^2, \qquad T = \Real\left(\frac{\cos%
\theta'}{\cos\theta}\sqrt{\frac{\varepsilon_2}{\varepsilon_1}}\,\right)%
\left|\frac{a_T}{a_I}\right|^2.
\end{equation}

Finding from the system (\ref{em_sys2}) the ratios $a_R/a_I$ and
$a_T/a_I$ and then substituting them into the equations (\ref{rt_cf1}),
one obtains the final equations for the $R$, $T$ and also for the
absorptance $A$ (see also \cite{LaYu2}):
\begin{eqnarray}
& & R = \left|\frac{U^{(1)}_P + U^{(2)}_P}{V^{(1)}_P + V^{(2)}_P}%
\right|^2, \label{r_cf}
\\
& & T = \Real\left(\frac{\cos\theta'}{\cos\theta}\sqrt{\frac%
{\varepsilon_2}{\varepsilon_1}}\,\right)\left|\frac{U^{(1)}_P %
V^{(2)}_P - U^{(2)}_P V^{(1)}_P}{V^{(1)}_P + V^{(2)}_P}\right|^2,
\label{t_cf}
\\
& & A = 1 - R - T. \label{a_cf}
\end{eqnarray}
If the two dielectric media are vacuum or air when $\varepsilon_1 = %
\varepsilon_2 = 1$, one has $\theta' = \theta$ and the equations
(\ref{r_cf}), (\ref{t_cf}) go over to the reflectance and transmittance
power coefficients presented in papers \cite{JoKlFu,KlFu2}.

\section{\sdt The surface impedance and the dielectric functions of the
degenerate electron plasma}

The surface impedance defined by equation (\ref{sf_imp1}), was evaluated
in \cite{JoKlFu} for the flat metallic film in the case of specular
electron reflecions from the film surface. For the P-wave it looks
as follows ($j = 1,2$):
\begin{equation}\label{sf_imp2}
Z^{(j)}_P = \frac{2\iim\Omega}{\beta W}\sum_n\frac{1}{Q^2_n}\left(%
\frac{Q^2_x}{\Omega^2\varepsilon_l(\Omega,Q_n)} + \frac{(\pi n/W)^2}%
{\Omega^2\varepsilon_{tr}(\Omega,Q_n) - (Q_n/\beta)^2}\right).
\end{equation}
Here $\Omega$, $\beta$, $W$, $Q_n$, $Q_x$ are dimensionless variables
and parameters:
\begin{eqnarray}
& & \Omega = \frac{\omega}{\omega_p}, \qquad \beta = \frac{v_F}{c},
\qquad W = \frac{\omega_p\,d}{v_F}, \label{ombtw}
\\
& & Q_n = \sqrt{\Bigl(\frac{\pi n}{W}\Bigr)^2 + Q^2_x}, \qquad
Q_x = \frac{v_F k_x}{\omega_p}. \label{qqx}
\end{eqnarray}
In (\ref{ombtw}) and (\ref{qqx}) $\omega_p$ is the degenerate electron
plasma frequency, $v_F$ is the electron Fermi velocity and the $k_x$
is the $x$-coordinate of the wave vector ${\bf k}$. Further, the
$\varepsilon_l(\Omega,Q)$ and $\varepsilon_{tr}(\Omega,Q)$ are the
respective longitudinal and transverse dielectric functions
(permittivities) of the electron gas. And summation in (\ref{sf_imp2})
is performed over all odd integers $n$ if $j = 1$ or over all even
integers if $j = 2$:
$$
\begin{array}{lll}
j = 1: & \ \ & n = \pm 1,\,\pm 3,\,\pm 5,\,\pm 7,\,\ldots;
\\
j = 2: & & n = 0,\,\pm 2,\,\pm 4,\,\pm 6,\,\ldots.
\end{array}
$$

It is convenient to use dimensionless variables in the units of the
degenerate electron plasma. The dielectric functions of the quantum
electron plasma at zero temperature with invariable relaxation time due
to electron collisions, obtained in the Mermin approach involving the
electron density matrix in the momentum space, look as follows
\cite{Mrm,LaYu5,LaYu6}:
\begin{eqnarray}
\varepsilon^{(qu)}_l(\Omega,Q) & = & 1 + \frac{3}{4 Q^2}\,\frac{%
(\Omega + \iim\gamma)F(\Omega + \iim\gamma,Q)F(0,Q)}{\Omega F(0,Q) +
\iim\gamma F(\Omega + \iim\gamma,Q)},
\label{epqu_l}
\\[.5em]
\varepsilon^{(qu)}_{tr}(\Omega,Q) & = & 1 - \frac{1}{\Omega^2}\left(1 +
\frac{\Omega G(\Omega + \iim\gamma, Q) + \iim\gamma G(0,Q)}%
{\Omega + \iim\gamma}\right).
\label{epqu_tr}
\end{eqnarray}
Here the functions $F$ and $G$ are defined by the equations:
\begin{eqnarray}
F(\Omega + \iim\gamma,Q) & = & \frac{1}{r}\Bigl[ B_1(\Omega_+ +
 \iim\gamma, Q) - B_1(\Omega_- +\iim\gamma, Q)\Bigr] + 2,
\nonumber\\[.5em]
G(\Omega + \iim\gamma,Q) & = & \frac{3}{16r}\Bigl[ B_2(\Omega_+ +
\iim\gamma, Q) - B_2(\Omega_- + \iim\gamma, Q)\Bigr] +
\nonumber\\
 & + & \frac{9}{8}\Bigl(\frac{\Omega + \iim\gamma}{Q}\Bigr)^2 +
\frac{3}{32}Q^2 r^2 - \frac{5}{8}, \nonumber
\end{eqnarray}
where the functions and variables
\begin{eqnarray}
& & B_1(\Omega + \iim\gamma, Q) = \frac{1}{Q^3}\bigl[(\Omega +
\iim\gamma)^2 - Q^2\bigr]\,L(\Omega + \iim\gamma, Q),
\nonumber \\
& & B_2(\Omega + \iim\gamma, Q) = \frac{1}{Q^5}\bigl[(\Omega +
\iim\gamma)^2 - Q^2\bigr]^2\,L(\Omega + \iim\gamma, Q),
\nonumber \\
& & L(\Omega + \iim\gamma, Q) = \ln\frac{\Omega + \iim\gamma - Q}%
{\Omega + \iim\gamma + Q}, \label{ln_rt}
\\
& & \Omega_{\pm} = \Omega \pm \frac{1}{2}Q^2 r. \nonumber
\end{eqnarray}
And further, the dimensionless variable and parameters
\begin{equation}\label{qgar}
Q = \frac{v_F |{\bf k}|}{\omega_p}, \qquad \gamma =
\frac{1}{\omega_p\tau}, \qquad r = \frac{\hbar\omega_p}{m_e v^2_F},
\end{equation}
where $\tau$ is the relaxation time owing to the electron collisions,
$m_e$ is the effective mass of the conductance electrons and $\hbar$ is
the Planck constant.

Some words about the logarithm ratio $L(\Omega + \iim\gamma, Q)$ defined
by (\ref{ln_rt}). Here the complex logarithm is defined on the complex
plane with the branching real negative half-line. And therefore, one has
to evaluate the logarithm ratio according to the rule:
\begin{equation}\label{lr_evl}
L(\Omega + \iim\gamma, Q) = \frac{1}{2}\ln\frac{(\Omega - Q)^2 + %
\gamma^2}{(\Omega + Q)^2 + \gamma^2} + \iim\left(\arctan\frac{\Omega %
+ Q}{\gamma} - \arctan\frac{\Omega - Q}{\gamma}\right).
\end{equation}
Using (\ref{lr_evl}) and taking into account that $Q\geqslant 0$ and
$\gamma\geqslant 0$ one can prove that the functions $F(0,Q)$ and
$G(0,Q)$ are real.

Note that the dielectric functions (\ref{epqu_l}), (\ref{epqu_tr}) in the
classical limit $r\to 0$ go over to corresponding dielectric functions
of the degenerate Fermi gas disregarding for the quantum wave electron
properties called as classical spatial dispersion case
\cite{KlFu1,LaYu2}:
\begin{eqnarray}
\varepsilon^{(cl)}_l(\Omega,Q) & = & 1 + \frac{3}{Q^2}\Biggl(1 +
\frac{\Omega + \iim\gamma}{2Q}L(\Omega + \iim\gamma, Q)\Biggr)\times
\nonumber
\\
 & \times & \Biggl(1 + \frac{\iim\gamma}{2Q}L(\Omega + \iim\gamma, Q)
\Biggr)^{-1}, \label{epcl_l}
\\[.5em]
\varepsilon^{(cl)}_{tr}(\Omega,Q) & = & 1 - \frac{3}{4\Omega}\Biggl(%
\frac{2(\Omega + \iim\gamma)}{Q^2} + B_1(\Omega + \iim\gamma, Q)\Biggr).
\label{epcl_tr}
\end{eqnarray}

And also, in the static limit $Q\to 0$ the quantum dielectric functions
(\ref{epqu_l}), (\ref{epqu_tr}) as well as the classical spatial
dispersion functions (\ref{epcl_l}), (\ref{epcl_tr}), go over to the
well-known classical Drude -- Lorentz electron dielectric function
without the spatial dispersion \cite{DrGr}:
\begin{equation}\label{epDL}
\varepsilon^{(DL)}_l(\Omega) = \varepsilon^{(DL)}_{tr}(\Omega)
= 1 - \frac{1}{\Omega(\Omega + \iim\gamma)}.
\end{equation}
Note that in the case of Drude -- Lorentz dielectric function
(\ref{epDL}), the summation in (\ref{sf_imp2}) can be done exactly.

Recall that the surface impedance (\ref{sf_imp2}) is used in (\ref{uv_c})
to evaluate the power coefficients (\ref{r_cf}) -- (\ref{a_cf}). In
(\ref{sf_imp2}), the $Q_x$ variable is evaluated by substituting $k_{1x}$
by (\ref{kx}) in the (\ref{qqx}) for $Q_x$ and taking into account the
definitions (\ref{ombtw}). As a result, one gets:
\begin{equation}\label{qx2}
Q_x = \Omega\beta\sqrt{\varepsilon_1}\sin\theta.
\end{equation}

\section{\sdt Numerical studies of power coefficients}

We performed numerical studies of the reflectance $R$, transmittance $T$
and absorptance $A$ power coefficients for P-wave. These coefficients
evaluated according to the equations (\ref{r_cf}) -- (\ref{a_cf}) with
use of (\ref{csthpr}), (\ref{uv_c}), (\ref{sf_imp2}), (\ref{qqx}) and
(\ref{qx2}). First, we studied the power coefficients evauated for the
quantum plasma with the dielectric functions (\ref{epqu_l}),
(\ref{epqu_tr}). Then we compared these coefficients with those
evaluated in the cases of the classical spatial dispersion approach
with dielectric functions (\ref{epcl_l}), (\ref{epcl_tr}) and in the
Drude -- Lorentz theory with the function (\ref{epDL}). And at the end,
we investigated the dependence of the power coefficients for the quantum
plasma on the parameter $r$ in (\ref{qgar}) called also as quantum
parameter.

We have taken the following data appropriate for potassium \cite{KlFu2}:
$v_F~=~8.5\cdot~10^5$~m/sec, $\omega_p~=~6.61\cdot~10^{15}$~sec$^{-1}$,
the mass of conductance electron equals to the mass of the free electron,
and also $\gamma = 10^{-3}$. Then using (\ref{ombtw}) and (\ref{qgar}),
we have adopted the dimensionless parameters $\beta = 2.83\cdot 10^{-3}$
and $r = 1.07$.

\begin{figure}[ht]

\vspace*{0mm}
\hspace*{-5mm}
\epsfig{file=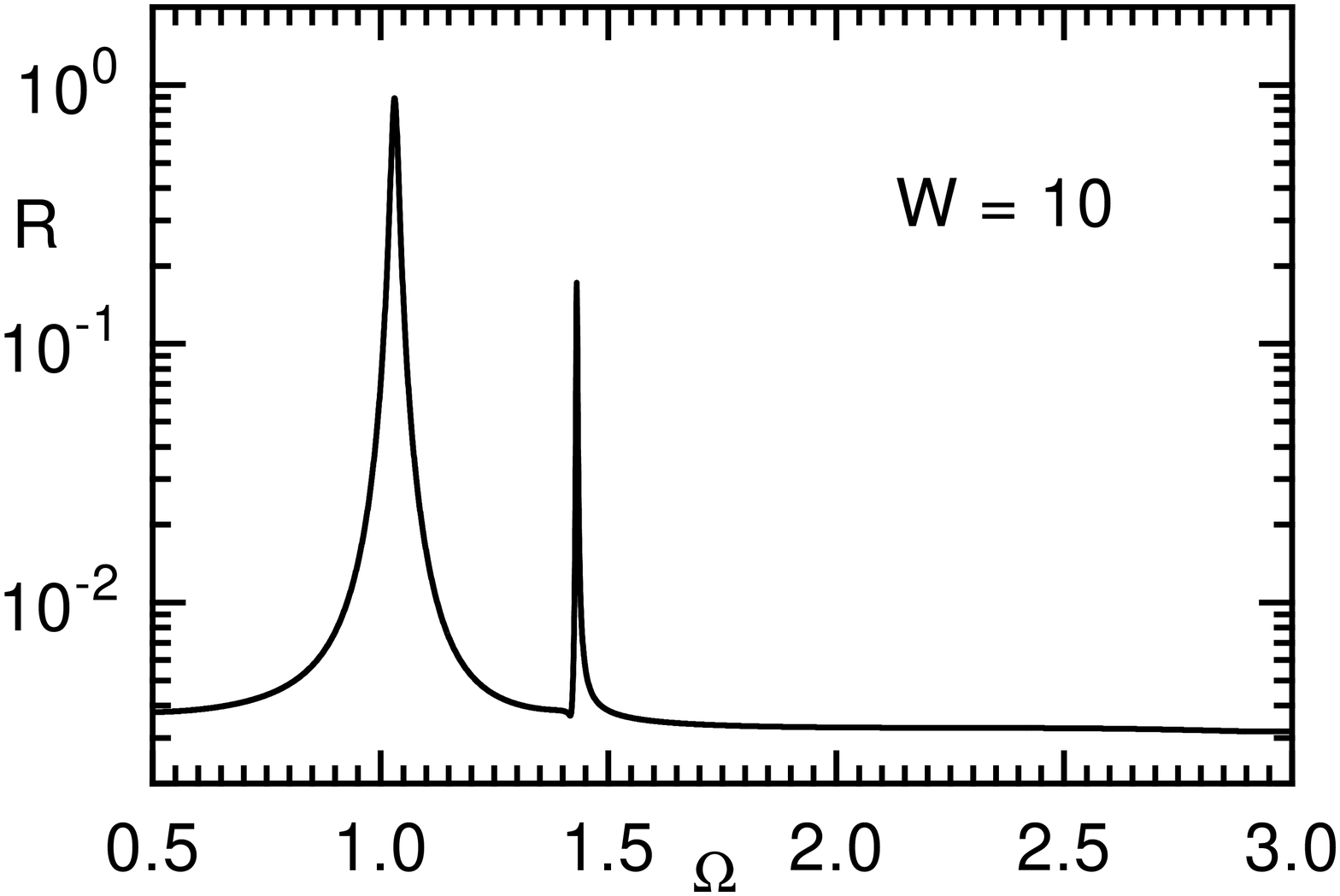,width=.53\textwidth}
\hspace*{-8mm}
\epsfig{file=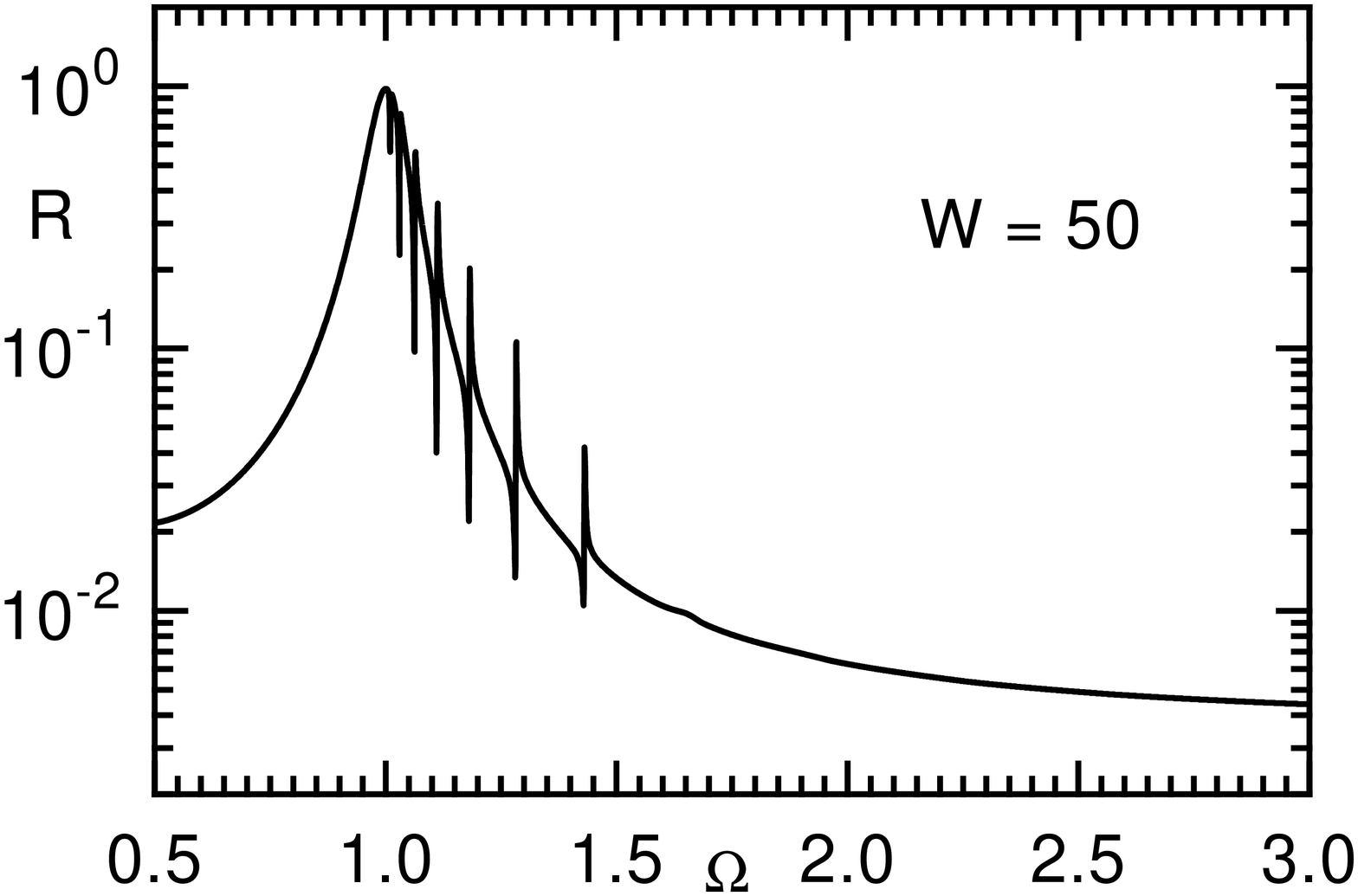,width=.53\textwidth}

\vspace*{0mm}

\caption{The reflectance $R$ as function of $\Omega$ for quantum
plasma at $\beta = 2.83\cdot 10^{-3}$, $\gamma = 10^{-3}$, $r = 1.07$,
$\theta = 60^{\circ}$, $\varepsilon_1 = 1$ (air), $\varepsilon_2 = 2$
(quartz), $W = 10$ (left plot) and $W = 50$ (right plot).}
\label{fig:r_qu}

\end{figure}

\begin{figure}[ht]

\vspace*{0mm}
\hspace*{0mm}
\epsfig{file=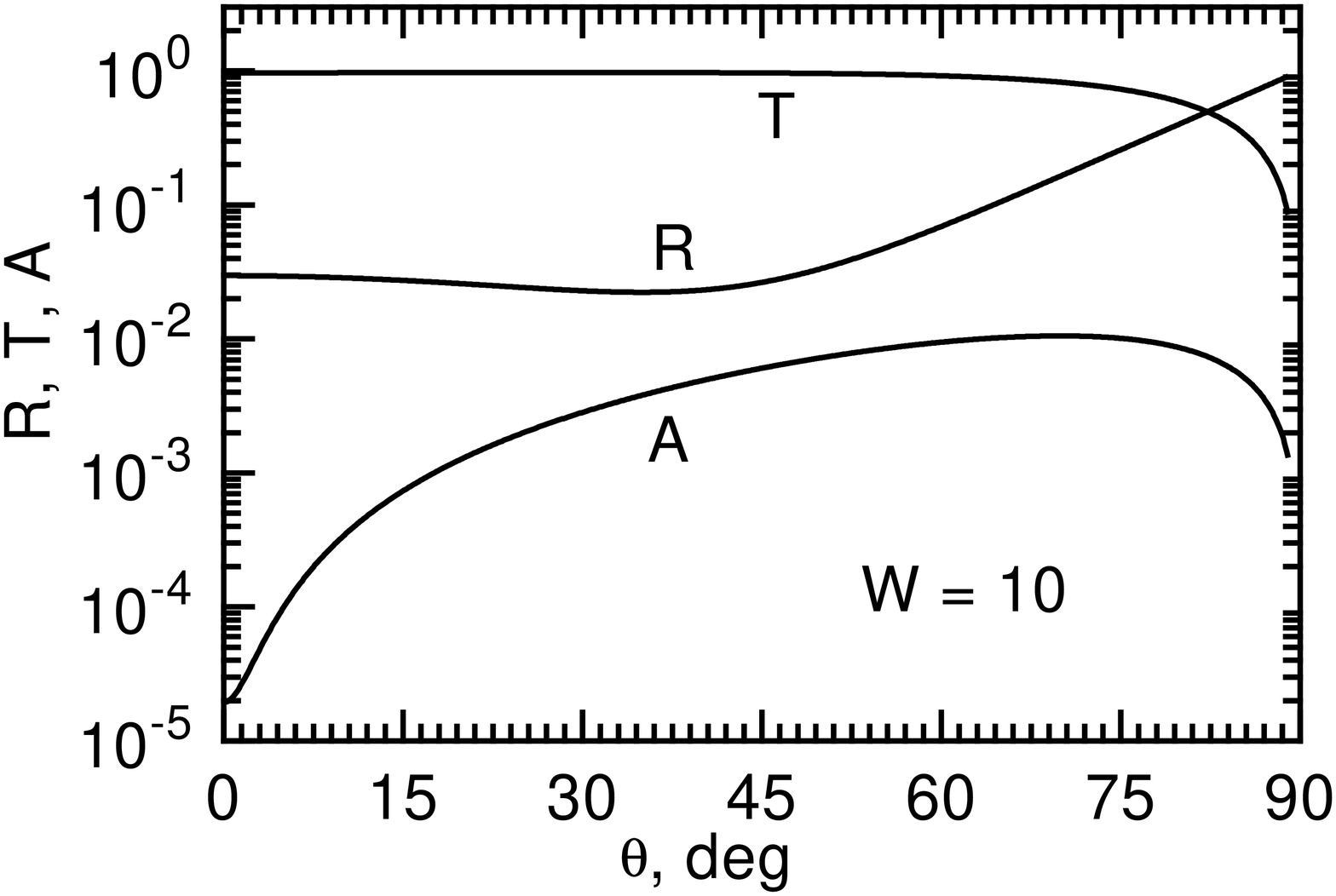,width=.5\textwidth}
\hspace*{-5mm}
\epsfig{file=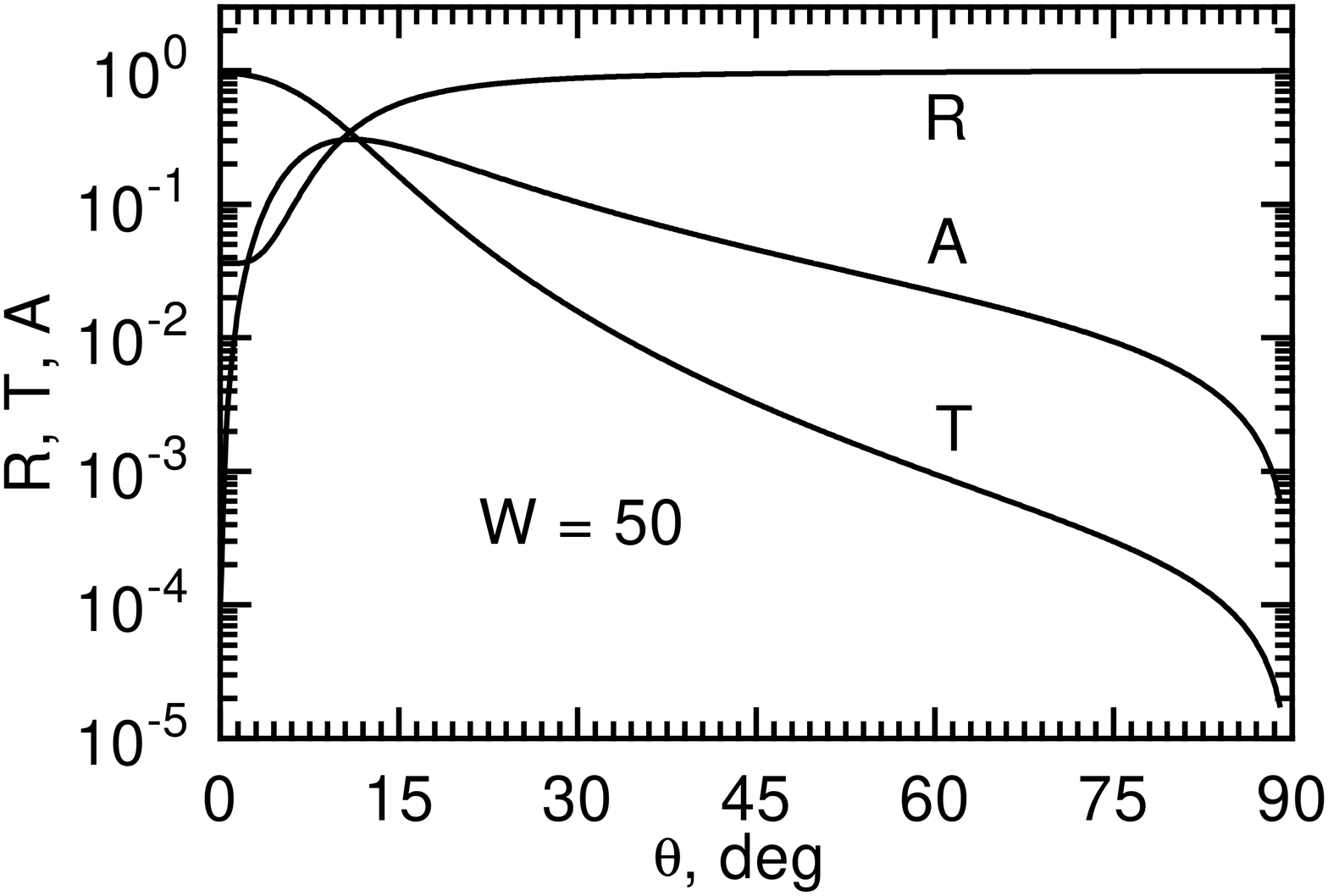,width=.5\textwidth}

\vspace*{0mm}

\caption{The coefficients $R$, $T$ and $A$ as functions of
$\theta$ for quantum plasma at $\beta = 2.83\cdot 10^{-3}$,
$\gamma = 10^{-3}$, $r = 1.07$, $\Omega = 1$, $\varepsilon_1 = 1$,
$\varepsilon_2 = 2$, $W = 10$ (left plot) and $W = 50$ (right plot).}
\label{fig:th_qu}

\end{figure}

At the fig.~\ref{fig:r_qu} we present numerical results for $R$ as
function of $\Omega$ at two values $W = 10$ and $W = 50$ corresponding
according to (\ref{ombtw}), to the film width $d = 1.286$~nm and
$d = 6.43$~nm respectively. The first dielectric medium taken by us is
an air with $\varepsilon_1 = 1$ and the second medium, or substrate, is
a quartz having $\varepsilon_2 = 2$. We considered the $\Omega$ in the
interval $0.5\leqslant\Omega\leqslant 3$ related to the visible and
ultraviolet light. One observes peaks near $\Omega = 1$ and when the
$\Omega > 1$. The distances between peaks decrease with growth of the
$W$. These results confirm the conclusions \cite{JoKlFu,PSChE} that the
peaks are caused by oscillations of longitudinal plasmons in the
metallic film and that the distances between peaks are of order $\pi/W$
or $\pi/d$.

Then we evaluate the power coefficients as functions of the incidence
angle $\theta$. Results are shown at the fig.~\ref{fig:th_qu} for the
same values $W$. One sees that the reflectance $R$ almost always
increases, the transmittance $T$ decreases with growth of the $\theta$
and the curves for $R$ and $T$ are intersecting. But the absorptance $A$
first have growth then it decreases. It is obvious also that the angle
of the $A$ maximum decreases with growth of $W$ or $d$ and this angle is
very close to the angle of the intersection of curves $R$ and $T$.

The above results for quantum electron plasma reproduce a qualitative
behaviour of the power coefficients for the classical degenerate spatial
dispersion plasma \cite{LaYu2}.

\begin{figure}[ht]

\vspace*{0mm}
\hspace*{-5mm}
\epsfig{file=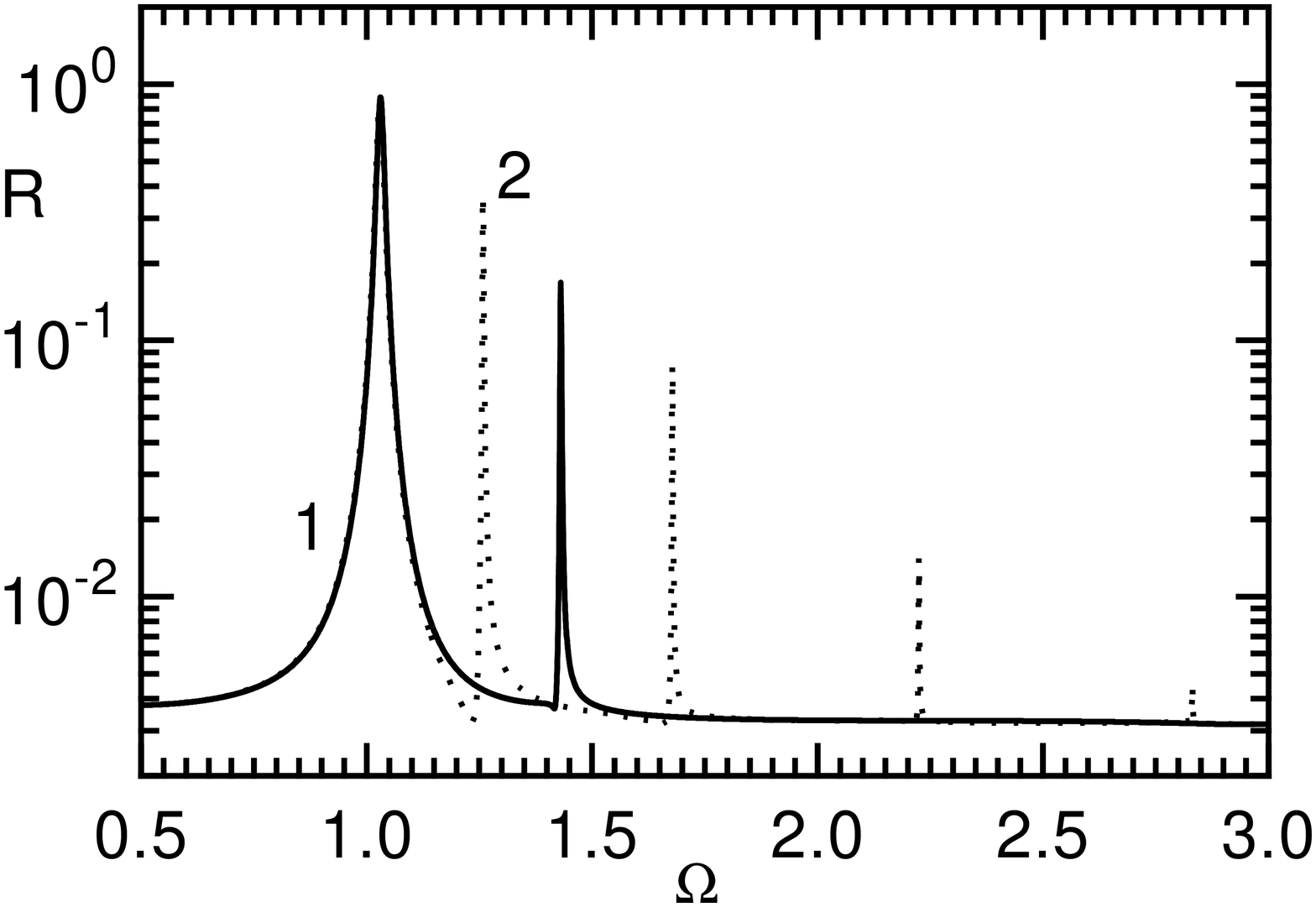,width=.53\textwidth}
\hspace*{-8mm}
\epsfig{file=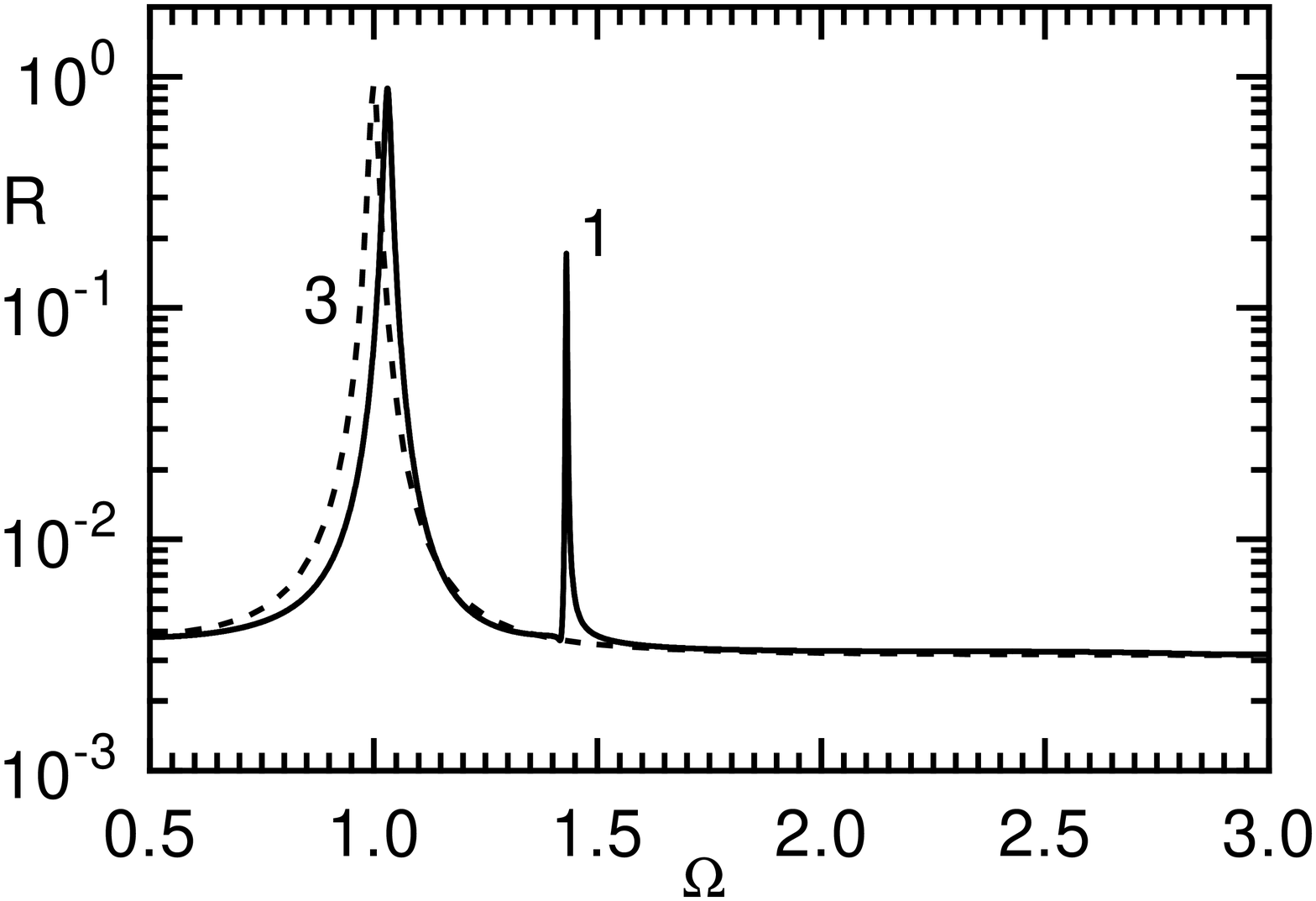,width=.53\textwidth}

\vspace*{0mm}
\hspace*{-5mm}
\epsfig{file=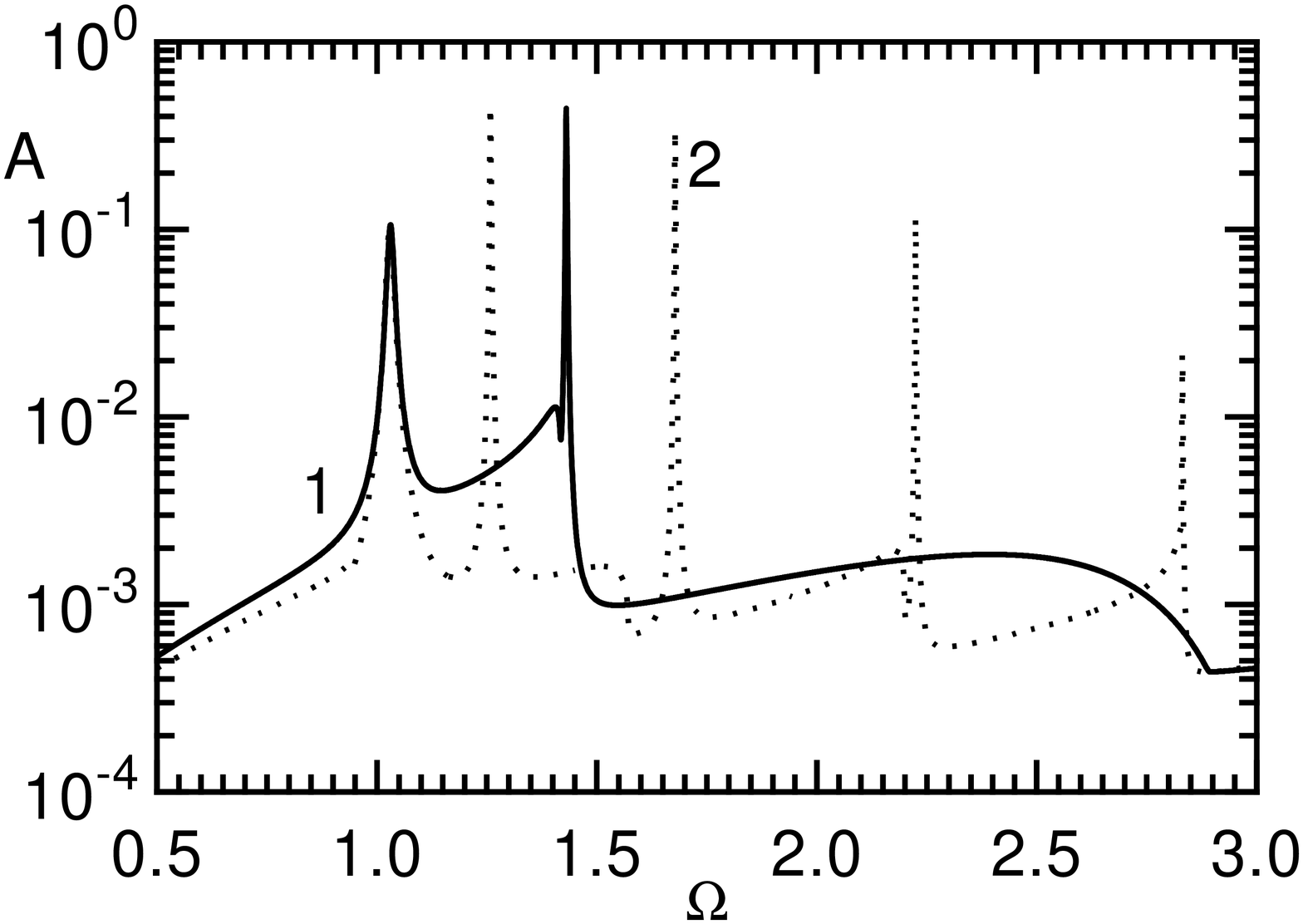,width=.53\textwidth}
\hspace*{-8mm}
\epsfig{file=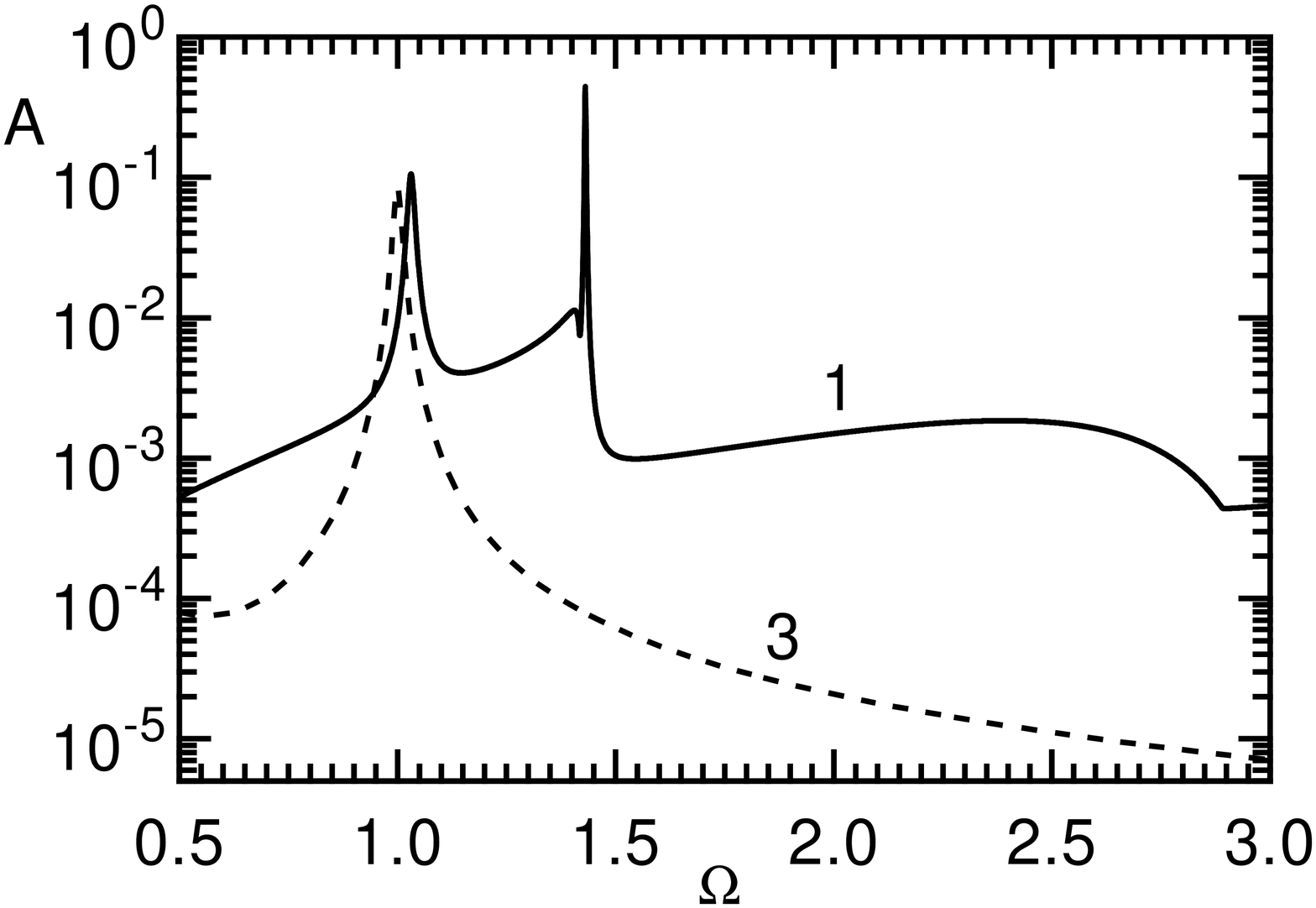,width=.53\textwidth}

\vspace*{0mm}

\caption{The reflectance $R$ (upper plots) and absorptance $A$
(lower plots) as functions of $\Omega$ at $\beta = 2.83\cdot 10^{-3}$,
$\gamma = 10^{-3}$, $r = 1.07$, $\theta = 60^{\circ}$,
$\varepsilon_1 = 1$, $\varepsilon_2 = 2$, $W = 10$: \ \ 1~--
quantum plasma (solid line), \ \ 2 -- classical spatial dispersion
approach (dotted line), \ \ 3 -- Drude -- Lorentz approach (dashed
line).}
\label{fig:ra_cf}

\end{figure}

\bigskip

Now we turn to a comparison of the quantum power coefficients with
those in the classical spatial dispersion and in the Drude -- Lorentz
approaches. Numerical results are presented at the fig.~\ref{fig:ra_cf}
for the reflectance $R$ and absorptance $A$ as functions of $\Omega$. We
took again the $\Omega$ in the interval $0.5\leqslant\Omega\leqslant 3$.

One sees that quantum results differ from the Drude -- Lorentz ones.
In the Drude -- Lorentz approach, the power coefficients have only
one peak in vicinity of $\Omega = 1$ related to the critical point of
the dielectric function (\ref{epDL}) whereas the quantum approach gives
peaks both near $\Omega = 1$ and at $\Omega \gtrsim 1$.

The power coefficients in the quantum approach differ also from those
in the classical spatial dispersion approach. As in the quantum case,
one has peaks in the classical theory. But the peaks in quantum and in
classical approaches are not coinciding at $\Omega > 1$. And quantity
of the classical peaks is greater than the quantum ones. So, one can
conclude that the quantum wave effects of electrons in the electron
plasma lead to a displacement, smoothing and vanishing of the most of
the resonant peaks.

\bigskip

\begin{figure}[ht]

\vspace*{0mm}
\hspace*{-5mm}
\epsfig{file=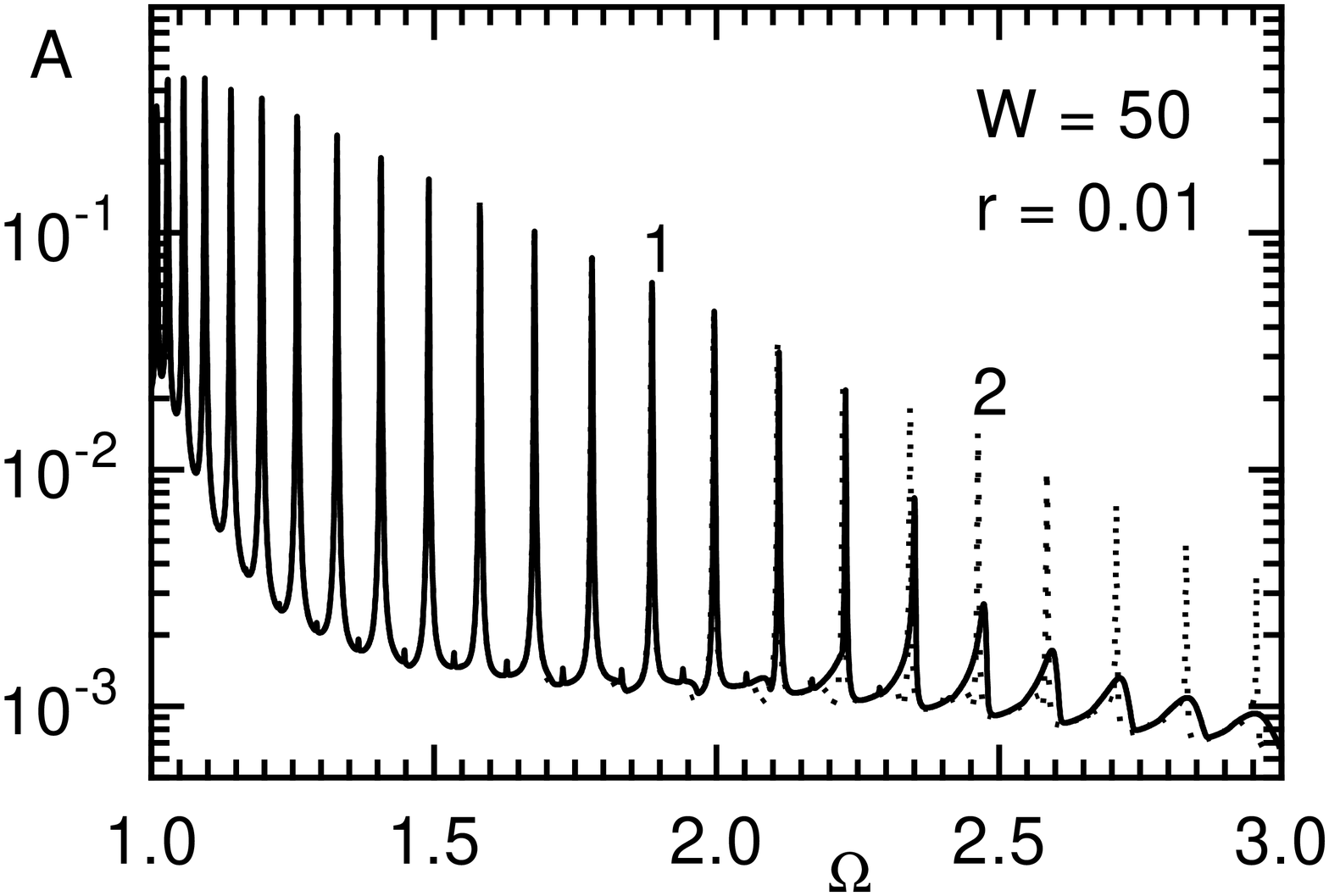,width=.53\textwidth}
\hspace*{-8mm}
\epsfig{file=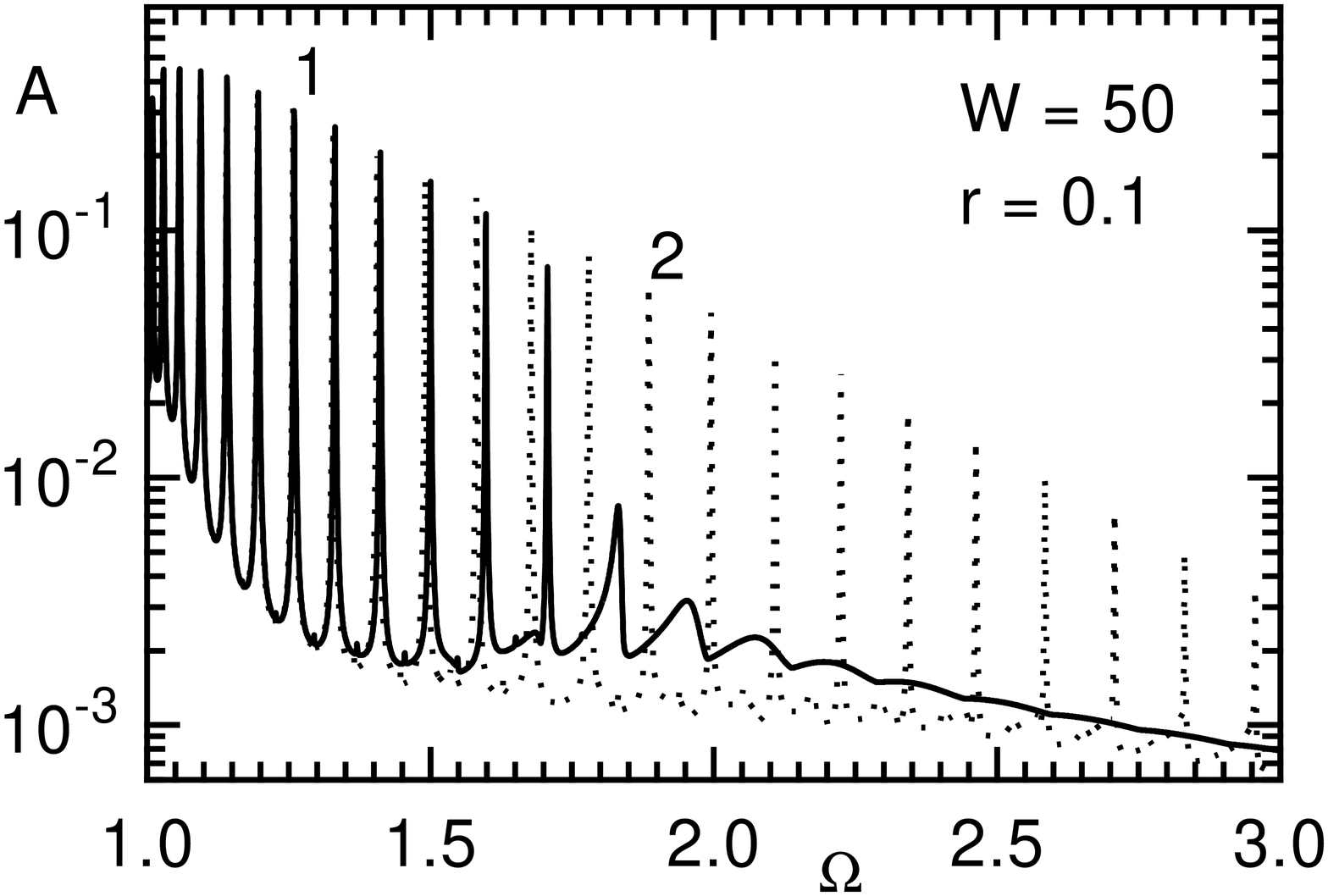,width=.53\textwidth}

\vspace*{0mm}
\hspace*{-5mm}
\epsfig{file=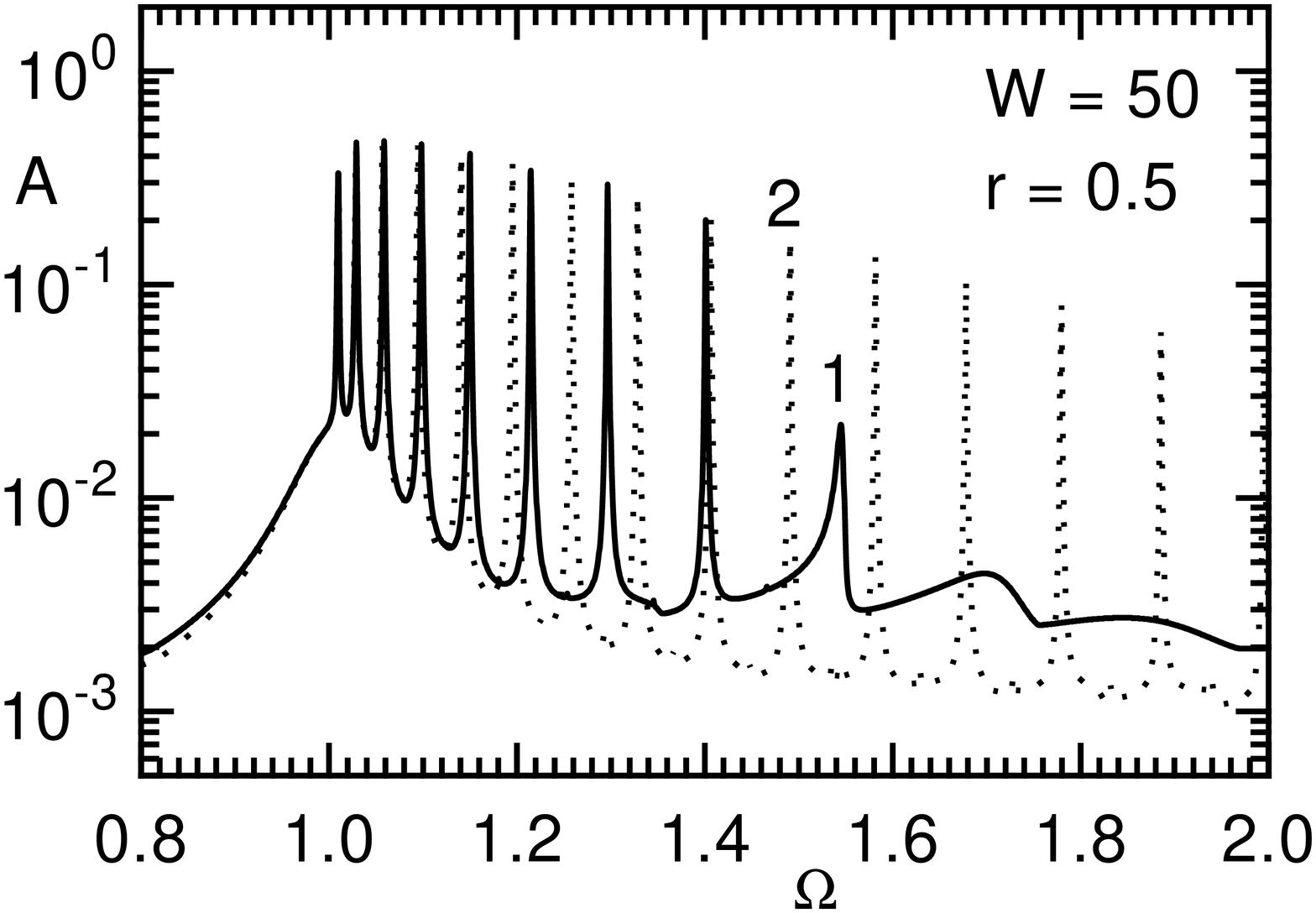,width=.53\textwidth}
\hspace*{-8mm}
\epsfig{file=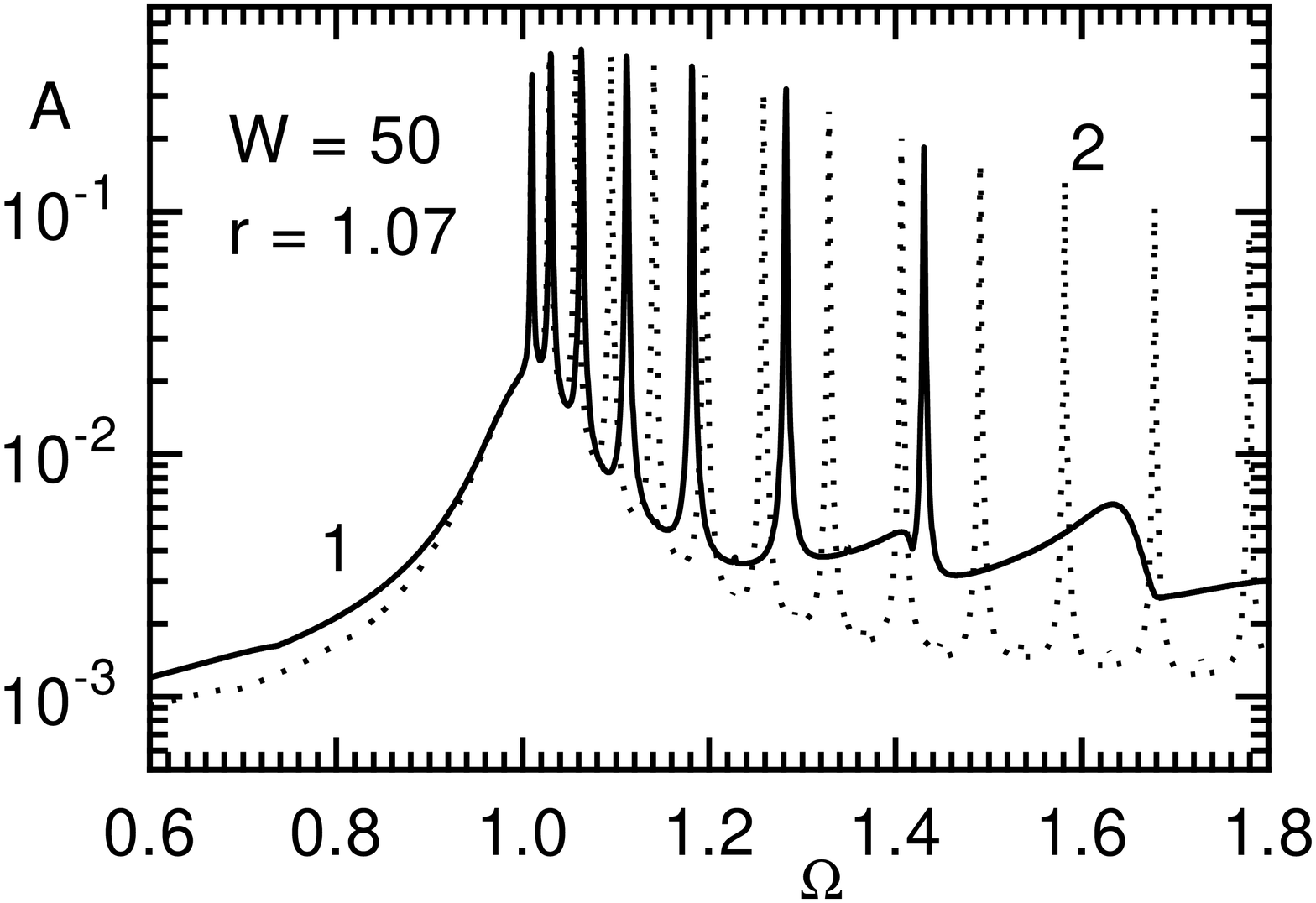,width=.53\textwidth}

\vspace*{0mm}

\caption{The absorptance $A$ as function of $\Omega$ at $\beta = %
2.83\cdot 10^{-3}$, $\gamma = 10^{-3}$, $\theta = 60^{\circ}$,
$\varepsilon_1 = 1$, $\varepsilon_2 = 2$, $W = 50$ for $r = 0.01$,
$0.1$, $0.5$ and $1.07$ values: \ \ 1 -- quantum plasma (solid line),
\ \ 2 -- classical spatial dispersion approach (dotted line).}
\label{fig:aqc_r}

\end{figure}

\begin{figure}[ht]

\vspace*{0mm}
\hspace*{0mm}
\epsfig{file=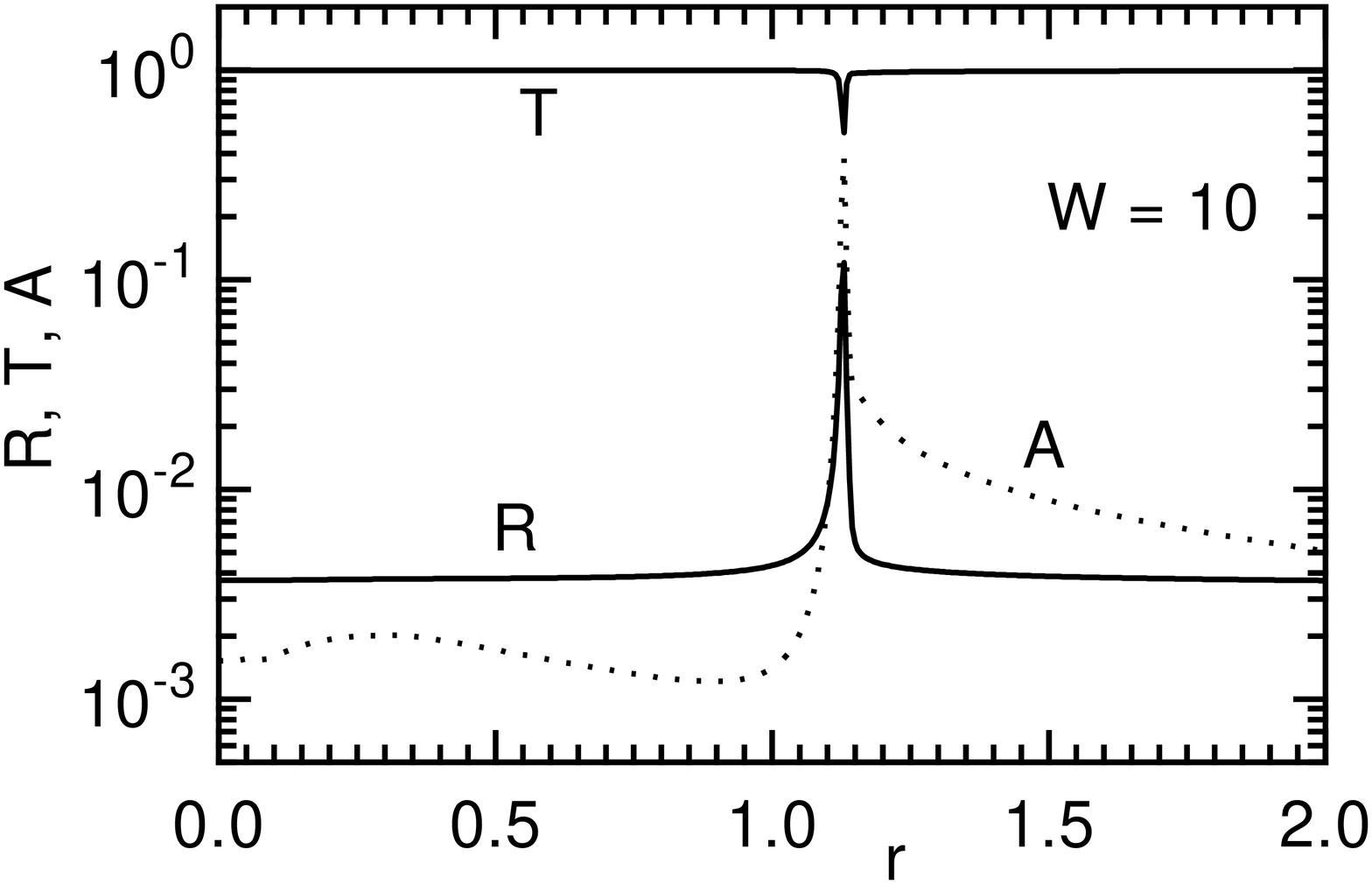,width=.5\textwidth}
\hspace*{-5mm}
\epsfig{file=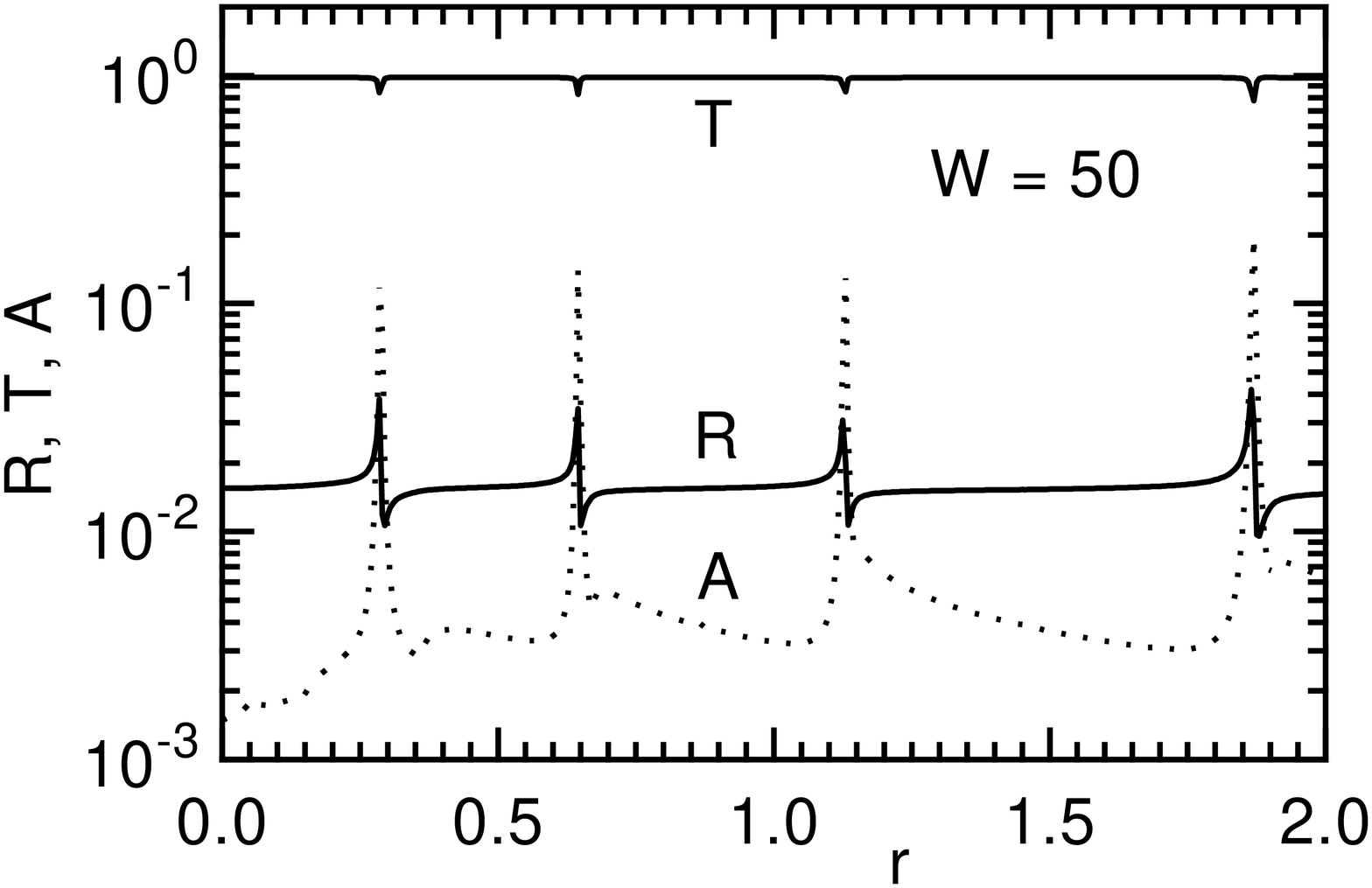,width=.5\textwidth}

\vspace*{0mm}

\caption{The coefficients $R$, $T$ (solid lines) and $A$ (dotted line)
as functions of $r$ for quantum plasma at $\beta = 2.83\cdot 10^{-3}$,
$\gamma = 10^{-3}$, $\theta = 60^{\circ}$, $\Omega = 1.45$,
$\varepsilon_1 = 1$, $\varepsilon_2 = 2$, $W = 10$ (left plot) and
$W = 50$ (right plot).}
\label{fig:cf_r}

\end{figure}

We made a more detailed investigation of the influence of the quantum
effects on the power coefficients. Typical results are presented at the
fig.~\ref{fig:aqc_r} and at the fig.~\ref{fig:cf_r}.

At the fig.~\ref{fig:aqc_r}, we show the quantum absorptance $A$ as
function of $\Omega$ for various small quantum parameter values $r$
in comparison with the one in the classical spatial dispersion approach
corresponding to the $r = 0$ case. One sees that as far as $r$ grows,
the difference between quantum and classical results becomes stronger.
And the disagreement starts at relatively large $\Omega$. Following
\cite{JoKlFu} and taking into account the equations (\ref{sf_imp2}),
(\ref{qqx}) with (\ref{epqu_l}) and (\ref{epqu_tr}), one can
qualitatively explain such a behaviour by an influence of the values
$rQ_n^2/2 \sim r(\pi n/W)^2$ on the dielectric functions for large
enough integers $n$. Hence the peaks for the $\Omega > 1$ caused by
contribution to the impedance (\ref{sf_imp2}) of terms with
$\Real\varepsilon_l(\Omega,Q_n) = 0$ in case of the odd integers $n$,
are shifted and disappear in quantum theory in comparison with the
classical spatial dispersion approach.

The dependences of the power coefficients on the quantum parameter $r$
are shown at the fig.~\ref{fig:cf_r}. One sees that the reflectance $R$
and the absorptance $A$ are much more sensitive to violation of the
parameter $r$ than the transmittance $T$. And at some $r$ values these
coefficients have a critical behaviour with peaks.

\section{\sdt Conclusion}

In this paper, we have studied numerically in the framework of the
quantum degenerate electron plasma theory the interaction of the
visible and ultraviolet P-waves with thin metallic film localized
between two dielectric media. We have chosen for the investigation
the reflectance, transmittance and absorptance power coefficients.
We have considered the dielectric functions of the quantum completely
degenerate electron plasma with invariable relaxation time in the
Mermin approach. Results for the quantum power coefficients were
compared with those both in the classical degenerate spatial dispersion
approach and in the Drude -- Lorentz one without spatial dispersion.

It has been detected that for the frequencies of order and larger than
the electron plasma frequency and related to the visible and ultraviolet
ranges, the power coefficients in quantum approach have the same
qualitative behavior as in the classical spatial dispersion degenerate
plasma theory. But the quantum power coefficients differ both from the
classical spatial dispersion and from the Drude -- Lorentz ones. The
quantum wave effects of electrons in electron plasma lead to
displasement, smoothing and vanishing of the classical degenerate
plasma peaks. Difference between quantum and classical approaches
becomes more visible at a growth of quantum parameter for relatively
large frequencies. The quantum power coefficients especially the
reflectance and absorptance, depend on the quantum parameter and these
dependences have some critical points.

The obtained results can be used in the theoretical investigations of
the quantum electron plasma as well as of the interaction of the
electromagnetic wave with thin metallic objects. These results may have
also practical applications for fine optical devices working with visible
and ultraviolet light.

\bigskip

This work is supported by the Research Grant of the President of Russian
Federation \ {\it MK-2382.2014.9} \ and also by the RFBR Grants \
{\it 15-37-20441 mol\_a\_ved} \ and \ {\it 14-47-03608} \ (jointly
with the Moscow Region Government).

\end{document}